\documentclass[twocolumn,showpacs,preprintnumbers,amsmath,amssymb,floatfix]{revtex4}

\usepackage{graphicx}
\usepackage{dcolumn}
\usepackage{bm}

\newcommand{\comment}[1]{}

\begin{document}

\title{Direct, Non-Destructive Imaging of
Magnetization in a Spin-1 Bose Gas}

\author{
J.\ M.\ Higbie, L. E. Sadler, S. Inouye, A. P. Chikkatur, S. R.
Leslie,\\ K. L. Moore, V. Savalli, and D.\ M.\ Stamper-Kurn}
\affiliation{Department of Physics, University of California,
Berkeley CA 94720}

\date{\today }

\begin{abstract}
Polarization-dependent phase-contrast imaging is used to spatially
resolve the magnetization of an optically trapped ultracold gas.
This probe is applied to Larmor precession of degenerate and
nondegenerate spin-1 $^{87}$Rb gases. Transverse magnetization of
the Bose-Einstein condensate persists for the condensate lifetime,
with a spatial response to magnetic field inhomogeneities
consistent with a mean-field model of interactions. Rotational
symmetry implies that the Larmor frequency of a spinor condensate
be density-independent, and thus suitable for precise magnetometry
with high spatial resolution. In comparison, the magnetization of
the noncondensed gas decoheres rapidly.
\end{abstract}


\maketitle

Quantum fluids with a spin degree of freedom have been of
longstanding interest, stimulated both by the complex
phenomenology of  superfluid   $^3$He \cite{voll90he3} and by
$p$-wave superconductivity \cite{mack03pwave}. Measurements of
magnetization and magnetic resonance have been crucial to
revealing the internal structure of these systems, inviting the
application of such techniques to related fluids. Advances in
ultracold atomic physics have now led to the creation of novel
multicomponent quantum fluids including pseudospin-$\frac{1}{2}$
Bose-Einstein condensates (BECs) \cite{hall98phas} and spin-1 and
-2 condensates of Na \cite{sten98spin,gorl03} and $^{87}$Rb
\cite{schm04,chan04,hira04}.

The internal state of a multi-component system is characterized by
the populations in each of the components and the coherences among
them.   However, in all previous studies of the spin-1 or spin-2
spinor condensates, while the populations in each magnetic
sublevel were measured, no information was obtained regarding the
coherence between overlapping populations
\cite{sten98spin,mies99meta,stam99tunprl,schm04,chan04,hira04}.
Moreover, although spatial patterns of longitudinal magnetization
have been reconstructed from images of freely expanding spinor
gases, the expansion process severely limits the resolution
obtainable \cite{mies99meta,lewa02,mcguirk02}.

\begin{figure}[htb]
    \mbox{
    \includegraphics[width=3in,viewport=0 40 270 270, clip]{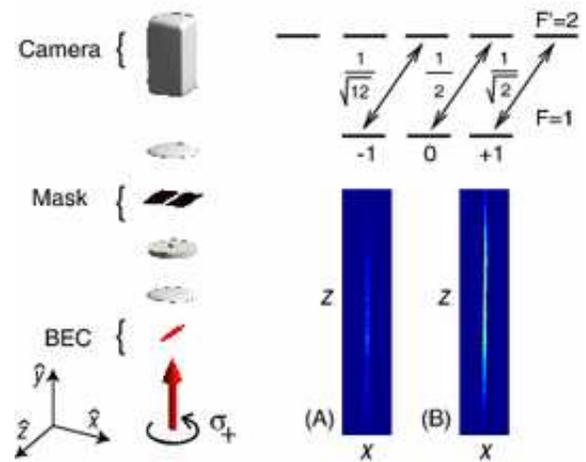}
    }

    \caption{\label{fig:scheme} Imaging system for direct
    detection of atomic magnetization.  Left: Circularly polarized probe
    light illuminates the trapped gas. A first lens and phase dot form a primary phase-contrast image which
    is selectively masked and then reimaged by a second lens
    onto the camera as one of $\sim40$ frames which form a single composite image.
    Top right: Clebsch-Gordan coefficients for the
    imaging transition. Bottom right: Sample images of a BEC (a) with
    the atomic spin along $-\hat{y}$ and (b) with the
    spin along $+\hat{y}$, demonstrating the magnetization sensitivity of our technique.
    {\it Higher-resolution version of figure at
    http://physics.berkeley.edu/research/ultracold }
    }
\end{figure}

In this work, we exploit atomic birefringence to image the
magnetization of an ultracold spin-1 Bose gas non-destructively
with high spatial resolution. By varying the orientation of an
applied magnetic field with respect to our imaging axis, we
measure either longitudinal magnetization, which derives from the
static populations in each of the magnetic sublevels, or
transverse magnetization, which derives from time-varying $\Delta
m\,$=$\,1$ coherences. This probe is used to observe Larmor
precession in both degenerate and non-degenerate spinor Bose
gases. In particular, optical characterization of Larmor
precession in a BEC provides a novel probe of the relative phases
between condensates in different internal states with excellent
temporal and spatial resolution (see Refs.\ \cite{recentphase} for
other recent measurements of condensate phase).


This work is related to experiments by the JILA group in which
either continuous \cite{matt99twist} or pulsed \cite{mcguirk02}
microwave fields were used to analyze a two-component
(pseudospin-$\frac{1}{2}$) $^{87}$Rb gas.  However, unlike this
psedospin system, the spin-1 spinor gas has more spin degrees of
freedom and hence a richer state space, the spin orientation is
connected directly with actual magnetization, and its internal
degree of freedom possesses vector symmetry, implying the rotation
invariance of its interactions. This rotational invariance
manifests itself in our work in the gapless nature of ``magnon''
excitations, and in the density independence of the
Larmor-precession frequency.

Our probe relies on the phase-contrast technique which permits
multiple-shot in-situ imaging of optically thick samples
\cite{andr97footnote}.  One can thus directly observe the dynamics
of a single gaseous sample, rather than reconstructing them from
experiments on many different samples. The phase-contrast imaging
signal strength for polarized probe light depends on both the
density and the internal state of the atoms being imaged. In the
case of imaging an $F=1$ gas of $^{87}$Rb with $\sigma_+$
circularly polarized light near the $F=1 \rightarrow F=2$ D1
transition, the phase-contrast signal is $\frac{1}{4}
\tilde{n}\sigma (\gamma / 2 \delta) \left(1+\frac{5}{ 6} \langle
F_y \rangle +\frac{1}{6}\langle F_y^2 \rangle\right)$ assuming
that the optical susceptibility of the dilute gas is small. Here
$\tilde{n}$ is the column number density,
$\sigma\,$=$\,3\lambda^2/{2\pi}$ the resonant  cross section,
$2\delta/\gamma$ the probe detuning in half linewidths, and $F_y$
the projection of the dimensionless atomic spin on the probe axis.
The phase-contrast signal is thus largely a local measure of one
vector component of the magnetization, when the density is known
\cite{caru04imag}.

We perform our experiments by collecting $5 \times 10^9$ $^{87}$Rb
Zeeman-slowed atoms in a magneto-optical trap, loading the cloud
into a Ioffe-Pritchard magnetic trap and evaporatively cooling it
to $\sim2\,\mu$K, before transferring the atoms into a
single-beam, linearly-polarized optical dipole trap (ODT). The ODT
derives from a free-running $825\,$nm diode laser, whose
fiber-coupled output is focused to diffraction-limited beam waists
of $(w_x, w_y)\,$=$\,(39, 13)\,\mu$m at the trap. For the
condensate studies, the ODT power is then ramped down over
$700\,$ms from $12\,$mW to a final value of $2.3\,$mW,
corresponding to trap frequencies
($\omega_x,\omega_y,\omega_z)\,$=$\,2\pi
(150,400,4)\,\mbox{s}^{-1}$ (axis orientations are indicated in
Fig.\ 1). The resulting evaporation yields nearly pure condensates
of  $4\times10^6$ atoms with a peak density of $5 \times 10^{14}
\, \rm{cm}^{-3}$. For the studies of the thermal cloud, we hold
the ODT power at $6.4\,$mW, yielding a gas of $\sim6 \times 10^6$
atoms at a temperature of $1.1 \, \mu$K, and a trap with
frequencies $(\omega_x,\omega_y,\omega_z)\,$=$\,2\pi (250,670,7)\,
\mbox{s}^{-1}$.

The atomic sample is phase-contrast imaged in two stages of
magnification ($\times 6$ and $\times 2$ sequentially) onto a CCD
camera. A physical mask blocks illumination of all but a narrow
slit-shaped region of the CCD chip. Taking advantage of a rapid
frame-shifting mode of our camera, we record 40 consecutive
images, each 25 pixels wide, at a rate of $20\,$kHz before
uploading the images.  We use probe light detuned $212\,$MHz below
the $F\,$=$\,1 \rightarrow F^\prime\,$=$\,2$ $D1$ transition
($\lambda\,$=$\,795\,$nm). Probe pulses are $5 \, \mu$s long with
an average intensity of $300 \,
\mu\rm{W}/\rm{cm}^2$\cite{acstarkfootnote}.

\begin{figure}[htb]
\mbox{
\includegraphics[scale=1,viewport=0 10 200 260,clip]{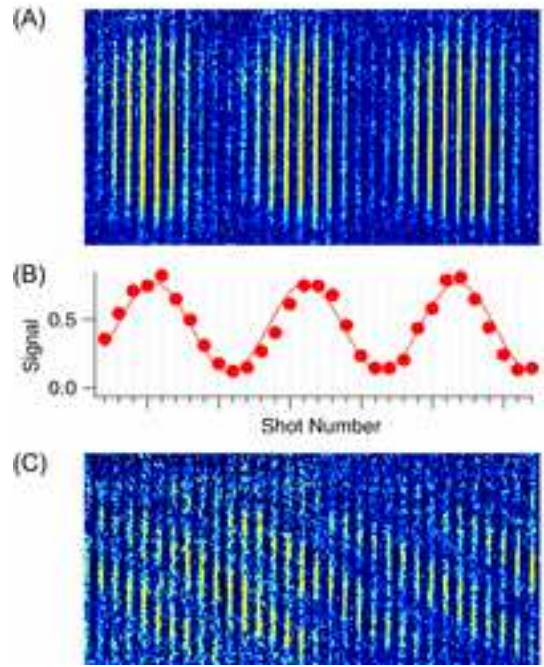}
} \caption{Direct imaging of Larmor precession of a spinor BEC
through magnetization-sensitive phase-contrast imaging. Shown are
31 consecutive images each with $325 \times 18 \,\mu$m field of
view. (a) Larmor precession is observed as a periodic modulation
in the intensities of repeated images of a single condensate. (b)
The peak signal strength oscillates at a rate which results from
aliased sampling of a precisely measured $38.097(15)\,$kHz Larmor
precession at a sampling rate of $20\,$kHz. (c) In the presence of
an $8\,$mG/cm axial gradient, images indicate ``winding'' of the
transverse magnetization along the condensate. Images begin (a)
$24.5\,$ms or (c) $14.5\,$ms after the tipping RF pulse. In (a),
the field gradient is cancelled to less than
$0.2\,$mG/cm.\label{fig:lp_img}
    {\it Higher-resolution version of figure at
    http://physics.berkeley.edu/research/ultracold }
    }
\end{figure}

We induce Larmor precession in our atomic sample starting with a
spin-polarized gas in the $|F\,$=$\,1,m_F\,$=$\,-1\rangle$ state
and a bias field of $54 \pm 2\,$mG along the $\hat{z}$ direction
\cite{bfieldfootnote}. A resonant RF pulse tips the spin vector by
an angle $\theta \simeq \pi/2$.  As the tipped spin precesses
about the bias field, the phase contrast image intensities
oscillate (Fig.\ 2a). We extract the peak signal from each of the
40 phase-contrast images by first binning  in the $\hat{z}$
direction over a small region ($27\,\mu$m for the BEC and
$108\,\mu$m for the thermal cloud) at the center of the cloud, and
then fitting to a $\mbox{sinc}$ function in the radial ($\hat{x}$)
direction to account for aberrations arising from imaging objects
near the $6 \, \mu$m imaging resolution limit. The temporal
oscillation in the peak height of our phase contrast images is
clearly visible (Fig.\ 2b), and is present only after the RF pulse
is applied. The transverse magnetization signals can be compared
to those from the static \emph{longitudinal} polarization of
spin-polarized samples in the $|F=1, m_F = \pm 1\rangle$ states
held in a magnetic field pointing along the imaging axis (Fig.\
1). This comparison confirms that the Larmor-precessing samples
are maximally polarized.  We note that the ratio of 4.2 observed
between the signals from spin-up and spin-down atoms is less than
the theoretical value of 6, perhaps because of imperfect probe
polarization.

The Larmor frequency of $\sim38\,$kHz is chosen in order to
maintain accurate control of the field direction and avoid
unwanted spin flips due to low-frequency field noise in our
laboratory.  Since twice the $20\,$kHz frame rate of our images is
within $\sim2\,$kHz of the Larmor frequency, we observe an aliased
low-frequency oscillation at the difference frequency. Combined
with the less precise, but absolute, determination of the RF pulse
resonance frequency, the Larmor precession signal measures in a
single shot the instantaneous magnetic field to a fraction of a
milligauss.


The amplitude of the Larmor precession signal gives a quantitative
measure of the coherence among Zeeman sublevels. Atomic gases with
long-lived and well-characterized coherences have important
scientific and technological applications. The lifetime of
magnetic-field-insensitive hyperfine coherences in ultracold atoms
has been studied \cite{lewa02,harb02pra,davi95}, with coherence
times up to $2$ s reported.  To measure the lifetime of
\emph{Zeeman} coherences in an ultracold Bose gas, we tip the
atomic spin using an RF pulse, as before, and then wait a variable
time before measuring the amplitude of the Larmor precession
signal (the ``tip and hold'' method). To isolate effects that
specifically diminish transverse magnetization from atom-number
loss or other systematic effects, we alternatively reverse the
order of the delay and the RF pulse (``hold and tip'').

\begin{figure}
\includegraphics[scale=.9]{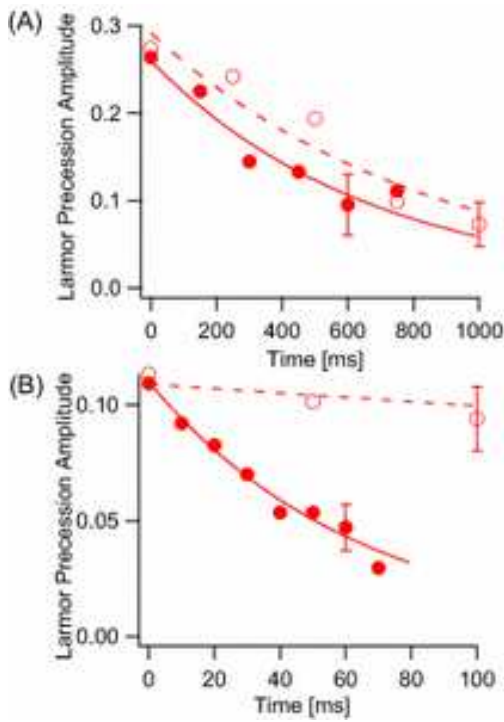}
 \caption{Decay of the Larmor precession amplitude for (a) a
BEC and (b) a thermal cloud. Data from the ``tip and hold''
(filled circles) or ``hold and tip'' (open circles) methods are
compared (see text). The  $1/e$ decay time of Larmor precession in
a BEC was $670 \pm 120\,$ms, close to that of the ``hold and tip''
signal ($830 \pm 120$\,ms), indicating no decoherence source other
than number loss. The $65 \pm 10\,$ms decay time of Larmor
precession in a thermal cloud was an order of magnitude shorter
than that the ``hold and tip'' signal ($1100 \pm
500\,$ms).\label{fig:LP_decay}}
\end{figure}

Results of such a measurement are shown in Fig.\ 3
\cite{binningfootnote}. The lifetime of transverse magnetization
in a spinor BEC, taken as the $1/e$ time of an exponential fit to
the precession amplitude vs.\ time, was $670\pm 120\,$ms, compared
to the measured $830\pm 120\,$ms condensate lifetime (determined
from the ``hold and tip'' method) in the optical trap. In other
words, no significant decoherence occurs other than that
attributable to overall number loss, consistent with 3-body decay
\cite{burt97}. We note that on other repetitions of the experiment
(data presented in Fig.\ 4), decoherence was observed on a
timescale of about 400\,ms, somewhat shorter than the condensate
lifetime. We cannot presently account for this  variation.

As the condensate is held for long times after the magnetization
is tipped into the transverse plane, the Larmor precession signal
begins to display a position-dependent phase shift (observed as
high as 25 radians) along the condensate axis, attributable to
magnetic field inhomogeneity (Fig.\ 2c).  For example, a gradient
of the field magnitude along the long axis of the condensate leads
to a ``corkscrew'' transverse magnetization, which winds up over
time (similar observations were made of pseudospin-$\frac{1}{2}$
spinor condensates \cite{matt99twist}). Qualitatively, the
condensed gas behaves as if its constituent atoms were frozen in
place, precessing at a frequency given by the local value of the
inhomogeneous magnetic field.

The response of the condensate magnetization to magnetic field
inhomogeneities can be described by the separate motion of each of
the three magnetic components present in a transversely polarized
cloud.  A magnetic field gradient imposes, over short times, a
$e^{i m q z}$ relative phase (or imparts a $m\hbar q$ relative
momentum) on the three components, labelled by the magnetic
quantum number $m$. This defines a ``corkscrew'' magnetization
with pitch $2 \pi /q$. At longer times, the inhomogeneous trapping
potential and condensate density begin to influence the motion of
the separate components, and hence the magnetization orientation.
For example, in a harmonically confined \emph{non-interacting}
spinor BEC, one would expect the following evolution in successive
quarter cycles of harmonic motion: magnetization winding up,
unwinding, winding up in the opposite sense, and unwinding again.
However, we observe no such dynamics, finding the phase of Larmor
precession at different portions of the condensate to advance
linearly with time through the $250\,$ms axial oscillation period.
This behavior, at least for the case of slight inhomogeneities,
can be explained by noting that slight rotations of the
magnetization orientation are associated with magnons possessing a
gapless and free-particle-like spectrum above the chemical
potential, which is constant across the condensate
\cite{hoandohmifootnote}. Thus, for an \emph{interacting} spinor
condensate, the combined effect of the inhomogeneous condensate
mean-field energy and the trap potential is to cause magnons, and
hence small-scale magnetization rotations, to advance as if no
external potential were present. We expect this argument to apply
similarly to larger variations of the magnetization orientation,
in accordance with our observations.

In contrast, transverse magnetization in a non-degenerate spinor
gas decays much faster than that in a BEC (Fig.\ 3). Moreover, the
Zeeman decoherence rate in a thermal gas depends strongly on the
applied magnetic field gradient (Fig.\ 4), while the (local)
Zeeman coherence in a BEC is unaffected by similar
inhomogeneities. For Fig.\ 4 we calibrated the axial field
gradient by fitting for the phase of the Larmor precession at
different axial positions in a condensate.  We can use this
information to cancel the axial gradient at the center of the trap
to better than $0.2\,$mG/cm, at which stage higher-order
inhomogeneities, most notably a field curvature of about
$20\,$mG/cm$^2$, dominate. More generally, this method may allow
for precise magnetometry with high spatial resolution ($\sim10 \,
\mu$m).

\begin{figure}
\mbox{
\includegraphics[scale=0.6,viewport=0 -15 300
250,clip]{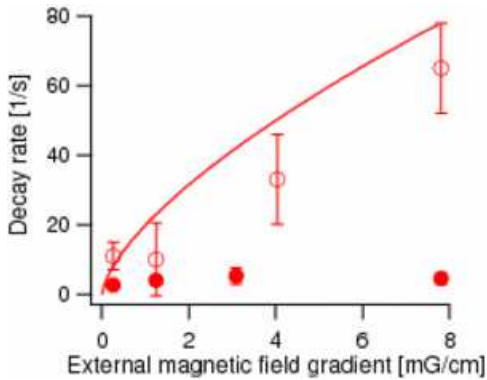} } \caption{\label{fig:grad_dep}
Dependence of Larmor precession decay rate on applied gradient.
Decay rates for the condensate are represented by filled circles
and for the thermal cloud by open circles. The solid curve is a
theoretical prediction for the thermal-cloud precession decay
rate, as discussed in the text.}
\end{figure}

The decay rates of transverse magnetization in a thermal cloud may
be estimated simply as $\Gamma_{\rm{LP}}=({\Omega'}_{\rm{L}}^2
v_{\rm{th}}/ \pi^2 n \sigma_c)^{1/3}$, where
$\Gamma_{\rm{LP}}^{-1}$ is the time for an atom to diffuse
\cite{diffusefootnote} into a field large enough to dephase its
spin by $\pi$ relative to a stationary spin, $v_{\rm{th}}$ is the
mean thermal velocity, $n$ the number density, $\sigma_c$ the
collisional cross-section, and
$2\hbar\Omega'_{\rm{L}}/\mu_{\rm{B}}$ the magnetic gradient. This
simple result is shown in Fig.\ 4 to agree fairly well with
measurement.  We note that this simple picture neglects spin
waves, which were observed in pseudospin-$\frac{1}{2}$ gases
\cite{bige89,lewa02,mcguirk02} and which should exist (in modified
form) in spin-1 Bose gases as well.

The demonstrated ability to image directly the magnetization of a
spinor Bose gas and the observation of long-lived Zeeman
coherences point to a number of future investigations. Given the
quadratic dependence of the dominant 3-body loss rate on
condensate density, much longer Zeeman coherence times may be
attained in lower-density spinor BECs. Furthermore, one expects
the Larmor precession frequency of a spinor BEC to be density
independent at low magnetic fields. This result stems from the
vectorial symmetry of the atomic spin, which requires the
zero-field interatomic interactions to be rotationally invariant
\cite{hoandohmifootnote}. Consequently, spinor BECs constitute an
attractive system for precise magnetometry.
Moreover, such magnetometry may be regarded as a form of
condensate-based interferometry, and as such invites similar
questions regarding the stability of the condensate phase, e.g.\
due to interactions \cite{grei02collapse} or reduced
dimensionality \cite{dett01phase}.  The fact that imaging the
transverse magnetization of a spin-1 gas resolves
\emph{spatially-varying} phase relations among \emph{three}
condensed components opens new experimental opportunities.

Our imaging method also promises to illuminate the mostly
unexplored properties of spinor condensates.  For example, spin-1
condensates of $^{87}$Rb are predicted to be ferromagnetic
\cite{hoandohmifootnote,klau01rbspin}. While measurements of
populations of the magnetic sublevels are consistent with this
prediction \cite{schm04,chan04}, no information on coherences, and
thus no conclusive evidence of ferromagnetism, has been obtained.
We are presently attempting to detect the spontaneous Larmor
precession of a $^{87}$Rb condensate as it relaxes to a
ferromagnetic ground state.  These experiments will be described
elsewhere.

We thank F.\ Lienhart, M.\ Pasienski, E.\ Crump and the UCB
Physics Machine and Electronics Shops for assistance, and W.\
Ketterle and D.\ Budker for helpful comments. This work was
supported by the NSF, the Hellman Faculty Fund, and the Alfred P.\
Sloan and the David and Lucile Packard Foundations. KLM
acknowledges support from the NSF and SRL from the NSERC.

\bibliographystyle{apsrev}
\bibliography{LP_bib,allrefs}

\begin{thebibliography}{28}
\bibitem[{acs()}]{acstarkfootnote}
\bibinfo{note}{Probe light at 300 $\mu$W/cm$^2$ imparts state-dependent AC
  Stark shifts equivalent to a $0.6\,$mG field along the probe direction. We
  observed no spin flips or number loss from our sample from this pulsed
  field.}
\bibitem[{bfi()}]{bfieldfootnote}
\bibinfo{note}{After nulling the magnetic field to $< 5$ mG, we apply
  additional fields either along or orthogonal to the imaging axis. Remaining
  field fluctuations cause shot-to-shot rms variations of 1.6 kHz in the Larmor
  precession frequency.}
\bibitem[{bin()}]{binningfootnote}
\bibinfo{note}{Here we fit the oscillation independently in each of three
  contiguous $27\,\mu$m binning regions and then average the three amplitudes
  obtained.}
\bibitem[{dif()}]{diffusefootnote}
\bibinfo{note}{Due to the large collision rate (200 Hz) compared with the axial
  oscillation frequency (7 Hz), thermal atoms move diffusively and slower than
  they would ballistically; also noted in Ref.\ \cite{harb02pra}.}
\end{thebibliography}

\end{document}